\documentclass[10pt]{elsart}
\usepackage{amssymb}
\usepackage{graphicx}
\usepackage{color}

\newcommand{\bea}{\begin{eqnarray}}
\newcommand{\eea}{\end{eqnarray}}
\newcommand{\bn}[1]{\mbox{\boldmath $#1$}}

\def\mb{\mbox}

\begin{document}

\begin{frontmatter}
\title{Interesting coupling phenomena of {\it heavy} and {\it ligth}
 holes in a $(GaAs/AlAs)^n$ superlattice\thanksref{talk}}

\author[UAM-A]{L. Diago-Cisneros\thanksref{peradd}}
\author[APFFUH]{H. Rodr\'{\i}guez-Coppola}
\author[TEFFUH]{R. P\'erez-\'Alvarez}
\author[UAM-A]{P. Pereyra}
\address[UAM-A]{Departamento de Ciencias B\'asicas, UAM-Azcapotzalco,
  C.P.02200 D.F. M\'exico.}
\address[APFFUH]{Depto. de F\'{\i}sica Aplicada, Fac. de F\'{\i}sica, UH,
  C.P.10400, La Habana,Cuba.}
\address[TEFFUH]{Depto. de F\'{\i}sica Te\'orica, Fac. de F\'{\i}sica, UH,
  C.P.10400, La Habana,Cuba.}
\thanks[talk]{Reduced version of a talk presented at the Madrid Meeting on Nanostructures
(Madrid, March 2003). Supported by CONACyT(Mexico) grand No.E120.1781, and UAM-A,
M\'exico, D.F.}
\thanks[peradd]{\textit{Permanent address}: Depto. de F\'{\i}sica Aplicada, Facultad de
               F\'{\i}sica, UH.}

\vspace*{-5mm}
\begin{abstract}
An appropriate combination of the scattering theory and the
transfer matrix formalism, for the solution of a $(4 \times 4)$
Kohn-L\"uttinger model, allow us to study the
multichannel-multiband transmission process of heavy and light
holes through a $(GaAs/AlAs)^n$ superlattice. Appealing effects
and interesting channel coupling phenomena, mediated by quasi-bond
states, are clearly foreseen.
\end{abstract}


\end{frontmatter}

\vspace*{-10mm}
\section{Theory: an outline of the Transfer Matrix-Scattering Matrix approach}
 \label{TM-SM}
\vspace*{-5mm} Following the theory of finite periodic systems \cite{PPP02}, we combine
the transfer matrix (TM) and the scattering matrix (SM) formalisms to describe, with
clear advantages and transparency, multichannel-multiband transmission processes. To
study the multichannel transport of heavy (\textit{hh}) and light (\textit{lh}) holes
through a semiconductor superlattice (SL) we consider a model in the framework of the $(4
\times 4)$ Kohn-L\"uttinger (KL) approach. At variance with the usual restrictions where
the incoming flux is a pure hole state \cite{Broido,Erdogan,WA,CC,LPRC02}, we have here
the possibility of a mixed-state input. Once the KL eigenvalue problem is solved, we
obtain the TM in the propagating mode representation and, hence, the $n$-cell
transmission amplitude\cite{PPP02}
\vspace*{-5mm}
\begin{equation}
\bn{t}_{n}
=\bn{\alpha}_{n}-\bn{\beta}_{n}\bn{\delta}_{n}^{-1}\bn{\gamma}_{n},
\end{equation}
\vspace*{-5mm}
with $\bn{\alpha}_{n}, \bn{\beta}_{n}, \bn{\delta}_{n}, \bn{\gamma}_{n}$,
$4\times 4$ ($n$-cell) TM blocks. Various transport coefficients can now be evaluated.
Among them, we will calculate the following global $n$-cell quantities:
\vspace*{-6mm}
\bea
 \label{TP&G}
T_{n_{ij}} = \left\vert t_{n_{ij}}
  \right\vert^{2};  \;\;\;\;
   & & T_{n_{i}}=  \sum_{j} \left\vert t_{n_{ij}}
  \right\vert^{2}  \;\;\;\mb{and}\;\;\;
  g_{n} = Tr \left(\bn{t}_{n}\bn{t}_{n}^{\dag}\right).
\eea
\vspace*{3mm} Here $i,j=hh_{+3/2}, lh_{-1/2}, lh_{+1/2}, hh_{-3/2}$ (ordered as in
Ref. \cite{Broido}), $T_{n_{ij}}$ is the transmission coefficient from channel $i$ to
channel $j$, $T_{n_i}$ the total transmission probability (through channel $i$), and
$g_n$ the two-probe superlattice Landauer Conductance (in units of $e^2/\pi \hbar$).

\section{Transmission Results}
 \label{Tran}
\vspace*{-5mm} We shall illustrate the just sketched approach with specific calculations
of \textit{hh} and \textit{lh} transmission coefficients through a $(GaAs/AlAs)^n$
superlattice, where the $GaAs$ well thickness will be $L_w = 50\;$\AA, the band off-set
$V_{b}= 0.55\;eV$ and the L\"uttinger parameters as in Ref. \cite{RKR}. For the $AlAs$
barrier thickness $L_b$ and for the number of cells $n$, we will take different values.
To keep an easy notation, we will drop the subindex $n$ from the physical quantities we
have plotted.

\vspace*{-3mm} In figure $1(a)$, ``direct path" transmission coefficients $T_{ii}$ and
the ``crossed path" $T_{ij}$ (with $i\neq j$), are plotted for the especial $n=1$ case,
and for transverse wave number $\kappa=0$. In figure $1(b)$ the transmission coefficients
$T_i$ (through the indicated channels $i$ ) and the conductance $g$ are plotted. The
conductance $g$ is shown also for other values of $\kappa$. For $\kappa=0$, all $T_{ij}$
but the crossed path $lh_{+1/2} \rightarrow hh_{-3/2}$ (dotted line) are
null\cite{LPRC02}. This result suggests some, low-level interference, and the passage of
flux from $lh_{+1/2}$ to $hh_{-3/2}$. This is congruous with the $T_i$ plots. There, the
$T_{lh_{1/2}}$ contribution remains around $0.5\;eV$. The $\kappa$ dependance of $g$ (see
r.h.s. panel), shows that increasing $\kappa$ the interaction and coupling between
channels (either opened or closed) gets enhanced. Since the conductance is a global
magnitude it does not reflect fine variations.

\vspace*{-3mm} In Figures $2$ we plot various transmission coefficients $T_{ij}$ for
$n=2$, and $i$ and $j$ as indicated in the right hand side panel. At low energies,
discrete quasi-stationary hole levels in the quantum well are correctly resolved
\cite{Erdogan,WA,RWS} ( levels are labelled as in reference [9]). At these energies (see
fig. $2(b)$), clear resonant transmission of flux from direct to crossed paths is seen.
Looking at the crossed paths $lh_{+1/2} \rightarrow hh_{-3/2}$ (full line) and $hh_{+3/2}
\rightarrow lh_{-1/2}$ (thick full line), for energies around $0.48\;eV$ and $0.56\;eV$,
interesting coupling phenomena between heavy- and light-hole propagating modes are also
apparent. As a consequence, it is possible that a hole entering this region as $hh_{\pm
3/2}$ or as $lh_{\pm 1/2}$ can exit in the same or a different state. Nevertheless, what
is more appealing is the fact that such resonant transport takes place without account of
the effective masses.

\vspace*{-3mm}
In figure $\ref{Fig4}$ we plot a direct path transmission probability for
different number of cells of the SL. The most visible effect is the formation of a
subband structure as $n$ grows \cite{PPP02}. In this case, the band spectrum reveals
itself reasonably well defined \cite{PPP02} when $n\approx5$. This number depends of
course on the SL parameters. As for 1D one channel periodic system, the band structure
can be strongly modified when these parameters are changed. To illustrate this, we plot
in figure $4$ the transmission probability for different values of the barrier thickness
$L_b$, while the remaining parameters are kept fixed. Reducing the barrier width, the
transmission coefficient grows (as it should) and, at the same time, the gaps become less
pronounced. Simultaneously the resonant peaks experience a kind of blue shift. It is
worth noticing however that, even though the band structure is a consequence of and will
emerge once the phase coherence and the periodicity have been combined, the single-cell
multichannel-multiband transfer matrix contains all the information of this fundamental
property \cite{PPP02,LPRC02}.
\begin{figure}
 \includegraphics[width=5.5in, height=2.5in]{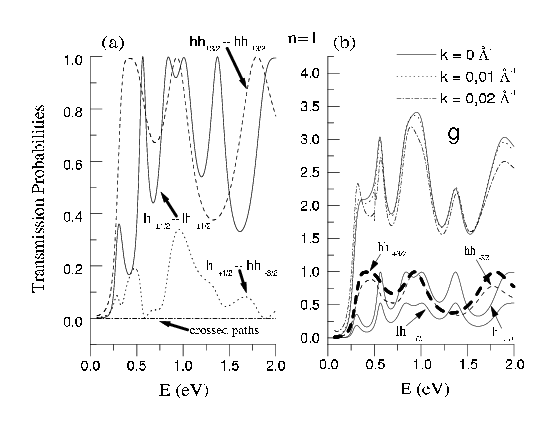}\\
 \caption{\label{Fig2} Transmission coefficients and conductance of
 $AlAs/GaAs/AlAs$, as functions of the
  incident particle energy. In (a), the Transmission coefficients $T_{ij}$ for $n=1$,
  $L_w+L_b=60\;$\AA $\,$ $\kappa=0$ and $i$ and $j$ as indicated. In
 (b), the total transmission coefficients $T_i$ through the indicated channel $i$, and
 the conductance $g$ for $\kappa= 0, 0.01$, and $ 0.02 \;$\AA$^{-1}$}
 \end{figure}
\begin{figure}
 \includegraphics[width=5.5in, height=2.5in]{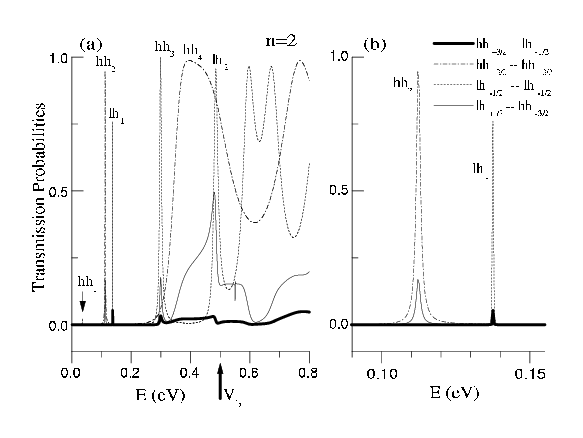}\\
 \caption{\label{Fig3} (a) The transmission coefficients $T_{ij}$ through
 a $(GaAs/AlAs)^n$ superlattice, with $n=2$ and total length of $120\;$\AA, as a function
 of the incident particle energy. At low energies the
discrete quasi-stationary hole levels are visible as isolated resonances. The height of
the barrier is pointed with an arrow in the energy axis. In (b) we have a zoom of the
lowest energy resonances. It is clear the channel coupling effect. The legend displayed
at the right hand panel is valid for both panels.}
\end{figure}
\begin{figure}
 \begin{center}
  \includegraphics[width=3in, height=2.5in]{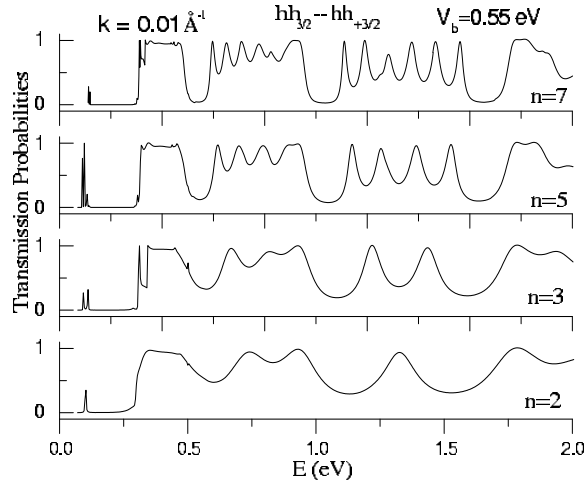}\\
 \end{center}
\caption{\label{Fig4} The Transmission probability for a
 ``direct" path $hh_{+3/2}\rightarrow hh_{+3/2}$ through a $(GaAs/AlAs)^n$
superlattice, as a function of the incoming particle energy, for different number of
cells $n$. Other parameters are $\kappa = 0.01\;$\AA$^{-1}$ and $V_b = 0.55$ eV. The
subband spectrum is visible when $n$ is of the order of $5$. The single cell length is
$60\AA$.}
\end{figure}
\begin{figure}
 \begin{center}
  \includegraphics[width=3in, height=2.5in]{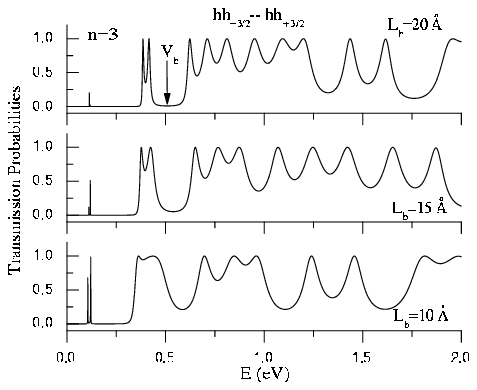}\\
 \end{center}
\caption{\label{Fig5} The same as Figure 3, for three barrier
thickness keeping fixed the well width at $\kappa=0$ for $n=3$. As
the length of the barrier is raised the structure of the spectrum
is better defined}
\end{figure}
\vspace*{-10mm}
\section{Summary}
 \label{Sum}
\vspace*{-5mm}
We have presented a more realistic TM-SM calculations for hole tunnelling
across different realizations of the $(GaAs/AlAs)^n$ superlattice. The system exhibits
clear channel interference effects between light- and heavy-hole quasiparticles. The
resonant transition from \textit{lh}-like modes to \textit{hh}-like modes is much
stronger than the other way around. Global quantities such as the conductance $g$ and
total transmission coefficient $T_i$, show that our model distinguishes pretty well the
four input-output channels, as well as, all the contributions to the tunnelling process.
Some of the features that we have presented are accessible only when one works in the
full $(4 \times 4)$ KL space within the proposed scheme of multichannel-multiband
transmission, of the $4$ available input channels.

\end{document}